\documentclass[10pt,twocolumn,showpacs,preprintnumbers,amsmath,amssymb,aps,prb,longbibliography,superscriptaddress,footinbib]{revtex4-2}
\usepackage{array,braket,mathtools,siunitx,bbm,enumitem,subcaption,bbold}

\usepackage[unicode=true,
bookmarks=true,bookmarksnumbered=true,bookmarksopen=false,
breaklinks=true,colorlinks=true]{hyperref}
\hypersetup{citecolor={blue},urlcolor={magenta}}
\AddToHook{cmd/appendix/before}{\crefalias{section}{appendix}}

\usepackage{cleveref}
\usepackage{caption}
\crefname{appendix}{App.}{Apps.}
\crefname{equation}{Eq.}{Eqs.}
\crefname{figure}{Fig.}{Figs.}
\crefname{table}{Tab.}{Tabs.}
\crefname{section}{Sec.}{Secs.}
\crefname{enumi}{Cond.}{Conds.}

\begin{document}
\title{Bootstrapping Disordered Quantum Systems}

\author{Yaprak Önder}
\thanks{These authors contributed equally to this work.}
\affiliation{Department of Physics, Harvard University, Cambridge, Massachusetts 02138, USA}

\author{Michael G. Scheer}
\thanks{These authors contributed equally to this work.}
\affiliation{Department of Physics, Harvard University, Cambridge, Massachusetts 02138, USA}

\author{Minjae Cho}
\affiliation{Leinweber Institute for Theoretical Physics, University of Chicago, Chicago, IL 60637, USA}

\author{Eslam Khalaf}
\email{eslam\_khalaf@fas.harvard.edu}
\affiliation{Department of Physics, Harvard University, Cambridge, Massachusetts 02138, USA}

\begin{abstract}
    We present a bootstrap framework that yields rigorous two-sided bounds on ground-state observables of quantum systems with quenched disorder. The key idea is to extend the operator algebra to include classical disorder variables and to encode the disorder distribution through a set of moments. The method directly bounds disorder-averaged expectation values for finite or infinite lattice systems and for arbitrary disorder distributions. In contrast to conventional approaches, we do not need to sample disorder realizations, which enables us to make thorough use of symmetries that hold only on average rather than for each disorder realization. Additionally, we show that multi-replica operators can be used to access disorder fluctuations such as the variance of an observable. We apply the method to the one-dimensional random transverse-field Ising model (RTFIM) with discrete, uniform, and Gaussian disorder, obtaining tight bounds on the disorder-averaged ground-state energy density and short-range spin-spin correlators, as well as bounds on the disorder variance of the energy density. Additionally, we show that the difference between the upper and lower bounds on the disorder-averaged ground-state energy density, which we term the bootstrap gap, can be used to map out the phase diagram of the RTFIM. The bootstrap gap is small and flat within the paramagnetic and ferromagnetic phases at weak disorder, grows upon entering the Griffiths regions, and peaks along the critical line.
\end{abstract}

\maketitle
\textit{Introduction.}---Quenched disorder gives rise to novel quantum many-body phenomena that are absent in clean systems, including Anderson and many-body localization \cite{Evers_2008,Nandkishore_2015}, infinite-randomness criticality \cite{PhysRevLett.69.534}, Griffiths effects \cite{Vojta_2006}, and quantum Hall plateaus \cite{AOKI1993951}. Despite their significance, disordered quantum many-body systems are difficult to study numerically. On top of the inherent difficulty of interacting problems, many approaches to disorder involve averaging over numerous disorder realizations, which can be very expensive \cite{Young_1996, Kr_mer_2024}. Furthermore, Griffiths (rare region) effects can give rise to slow equilibration of simulations near criticality \cite{Hukushima_2008}. These challenges motivate the search for alternative numerical techniques capable of directly accessing disorder averaged observables. 

The bootstrap approach is a powerful computational method with wide-ranging applications in high energy theory, quantum chemistry, condensed matter theory, and quantum information \cite{Han_2020,Lin_2020,Kazakov_2023,Navascu_s_2007,Mazziotti_2023,berenstein2024numericalexplorationbootstrapspin,Wang_2024,schouten2025bootstrappingelectronicstructurequantum,Fawzi_2024,Cho:2024kxn,Cho:2024owx,Cho:2025vws,csnn-vjhn,paul2026bootstrappinggroundstateproperties,Gao_2026,Scheer_2026,scheer2025defectbootstraptightground,chadha2026bootstrapboundsquantumspin}. In the condensed matter context, the bootstrap method yields rigorous, two-sided bounds on expectation values in ground states or thermal states of classical and quantum Hamiltonian systems \cite{Fawzi_2024,Cho:2024kxn,Cho:2024owx,Cho:2025vws,csnn-vjhn,paul2026bootstrappinggroundstateproperties,Gao_2026,Scheer_2026,scheer2025defectbootstraptightground,chadha2026bootstrapboundsquantumspin}. Previous studies have focused on clean systems, and the possibility of including disorder in bootstrap has so far remained unexplored. In this letter, we present a bootstrap framework for disordered quantum systems. By extending the ordinary operator algebra to include classical disorder variables with prescribed probability distributions, we derive bootstrap bounds on disorder-averaged expectation values that hold rigorously on finite or infinite lattices. The method does not require sampling disorder realizations and is applicable to arbitrary disorder distributions. Furthermore, since we are accessing disorder-averaged quantities directly, our formalism allows us to exploit symmetries which hold only on average. This is often a much larger symmetry group than the one that holds for every disorder realization. We note that a similar idea was employed to incorporate disorder in tensor network simulations \cite{CiracDisorderPRL, CiracDisorderSciPost, BultinckDisorder}.

As an illustrative example, we apply our method to the 1D random transverse-field Ising model (RTFIM), which is a nearest-neighbor Ising model with a random transverse field at each site. This system exhibits a paramagnet-to-ferromagnet phase transition characterized by an infinite-randomness fixed point and additionally exhibits Griffiths (rare region) effects \cite{PhysRevLett.69.534,PhysRevB.51.6411}. The clean case (the 1D transverse field Ising model) has recently been investigated using the bootstrap framework \cite{Fawzi_2024,Fawzi_2024,Cho:2024owx,scheer2025defectbootstraptightground,chadha2026bootstrapboundsquantumspin}. We obtain bounds on the disorder-averaged ground-state energy density and spin-spin correlators, as well as on the variance of the ground-state energy density with respect to disorder. Additionally, we show that the difference between the upper and lower bounds on the energy, hereafter referred to as the bootstrap gap, serves as a useful probe of the phase diagram of the RTFIM. The bootstrap gap is small in the paramagnetic and ferromagnetic phases with weak disorder, starts increasing near the onset of the Griffiths regions, and exhibits a peak centered at the critical line. This is consistent with the behavior observed in bootstrap bounds for the classical Ising model \cite{Cho:2022lcj} and clean quantum systems \cite{scheer2025defectbootstraptightground,chadha2026bootstrapboundsquantumspin}, where the difference between upper and lower bounds on the ground-state energy density can serve as an indicator for phase boundaries. We also report a similar bootstrap formulation for classical equilibrium and nonequilibrium systems with quenched disorder in a parallel work \cite{unpubl1}.

\textit{Setup.}---We begin by describing the formalism for the case of a finite dimensional quantum system with quenched disorder. We denote by $\mathcal{L}_{\text{phys}}$ and $\mathcal{B}_{\text{phys}}$ the spaces of all operators and Hermitian operators on the physical Hilbert space $\mathcal{H}_{\text{phys}}$, respectively. The set of disorder realizations is a probability space $\Omega$ with measure $\mu$. For simplicity, we assume for now that $\Omega$ is finite, so there are finitely many disorder realizations. For some real-valued random variables (RVs) $J_\alpha : \Omega \to \mathbb{R}$ and Hermitian operators $H_\alpha \in \mathcal{B}_{\text{phys}}$, the disordered Hamiltonian $H : \Omega \to \mathcal{B}_{\text{phys}}$ is a Hermitian-valued RV $H = \sum_\alpha J_\alpha H_\alpha$.

Our goal is to bound disorder-averaged ground-state properties of $H$. That is to say, let $\rho : \Omega \to \mathcal{B}_{\text{phys}}$ be a Hermitian-valued RV such that $\rho(\omega)$ is a ground state density matrix of $H(\omega)$ for each disorder realization $\omega \in \Omega$. We would like to bound quantities such as the disorder-averaged ground-state energy
\begin{equation}\label{eq:disordered-averaged-energy}
\sum_{\omega \in \Omega} \text{tr}(H(\omega) \rho(\omega)) \mu(\omega).
\end{equation}

The key idea is to promote the probability space $\Omega$ to a Hilbert space $\mathcal{H}_\Omega$ with orthonormal basis $\ket{\omega}$ for each $\omega \in \Omega$. We denote by $\mathcal{L}_\Omega$ and $\mathcal{B}_\Omega$ the spaces of all operators and Hermitian operators $\mathcal{H}_\Omega$, respectively. We now define the unquenching map $u$ for a real-valued RV $A$, operator-valued RV $B$, or Hermitian operator $C \in \mathcal{B}_{\text{phys}}$ by
\begin{gather}
u(A) = \sum_{\omega \in \Omega} A(\omega) \ket{\omega} \bra{\omega} \otimes \mathbb{1}_{\text{phys}} \nonumber \\
u(B) = \sum_{\omega \in \Omega} \ket{\omega} \bra{\omega} \otimes B(\omega)\qquad
u(C) = \mathbb{1}_\Omega \otimes C.
\end{gather}
Here, $\mathbb{1}_\Omega \in \mathcal{B}_\Omega$ and $\mathbb{1}_{\text{phys}} \in \mathcal{B}_{\text{phys}}$ denote identity operators so that $u(X) \in \mathcal{L}_\Omega \otimes \mathcal{L}_{\text{phys}}$ for all $X$. If we additionally define the disorder-averaged ground-state
\begin{equation}
\rho_\mu = \sum_{\omega\in \Omega} \mu(\omega) \ket{\omega}\bra{\omega} \otimes \rho(\omega)
\end{equation}
which is an element of $\mathcal{B}_\Omega \otimes \mathcal{B}_{\text{phys}}$, then $\text{tr}(u(X) \rho_\mu)$ is the disorder-averaged ground state expectation of $X$. For example, $\text{tr}(u(H) \rho_\mu)$ is precisely the disorder-averaged ground-state energy in \cref{eq:disordered-averaged-energy}.

In order to bootstrap disorder-averaged ground-state expectations, we consider what properties any disorder-averaged ground state $\rho_\mu$ must satisfy. Since $\text{tr}(\rho(\omega)) = 1$ for each $\omega$, we have
\begin{equation}\label{eq:moments}
\text{tr}(u(R) \rho_\mu) = \sum_{\omega \in \Omega} \mu(\omega) R(\omega)
\end{equation}
for any real-valued RV $R$. Since $\rho(\omega) \succeq 0$ for each $\omega$, it satisfies the positivity condition
\begin{equation}\label{eq:positivity}
\text{tr}(u(\mathcal{O})^\dagger u(\mathcal{O}) \rho_\mu) \geq 0
\end{equation}
for any operator-valued RV $\mathcal{O}$. Finally, since $\rho(\omega)$ is a ground state of $H(\omega)$ for each $\omega$, it satisfies the perturbative positivity condition
\begin{equation}\label{eq:perturbative-positivity}
\text{tr}(u(\mathcal{O})^\dagger [u(H), u(\mathcal{O})] \rho_\mu) \geq 0
\end{equation}
for any operator-valued RV $\mathcal{O}$.

We now consider the generalization of this construction to replicated systems. For simplicity, we focus here on the two-replica case, which can be used to bound the variance of observables with respect to disorder. For each disorder realization $\omega$ we are interested in a $\rho^{(2)}(\omega) = \rho(\omega) \otimes \rho(\omega)$ which is a ground state of $H^{(2)}(\omega)$, where
\begin{equation}
H^{(2)} = H \otimes \mathbb{1}_{\text{phys}} + \mathbb{1}_{\text{phys}} \otimes H.
\end{equation}
$H^{(2)}$ is a Hermitian-valued RV acting on the space $\mathcal{H}^{(2)}_{\text{phys}} = \mathcal{H}_{\text{phys}} \otimes \mathcal{H}_{\text{phys}}$. In order to include disorder, we simply repeat the above construction with the replacements $\mathcal{H}_{\text{phys}} \mapsto \mathcal{H}^{(2)}_{\text{phys}}$, $\rho \mapsto \rho^{(2)}$, and $H \mapsto H^{(2)}$. There is, however, one new positivity condition that we can add. For any operator-valued RV $\mathcal{O}$, we note that
\begin{equation}
\text{tr}((\mathcal{O}^\dagger \otimes \mathcal{O}) \rho^{(2)}(\omega)) = |\text{tr}(\mathcal{O} \rho(\omega))|^2 \geq 0.
\end{equation}
We therefore must have
\begin{equation}
\text{tr}(u(\mathcal{O}^\dagger \otimes \mathcal{O}) \rho^{(2)}_\mu) \geq 0 
\end{equation}
for any operator-valued RV $\mathcal{O}$. We note that this is the disorder generalization of the PPT criterion for replica-swap symmetric states \cite{Peres1996,HORODECKI19961,T_th_2009}.

So far, we have discussed the case of a finite-dimensional Hilbert space and a finite probability space. However, one can use the algebraic quantum mechanics formalism to generalize this to local infinite systems and arbitrary probability spaces \cite{bratteli2003operator}. In this case, $\mathcal{L}_\Omega \otimes \mathcal{L}_{\text{phys}}$ is generalized to an algebra $\mathcal{A} = \mathcal{A}_\Omega \otimes \mathcal{A}_{\text{phys}}$ of quasi-local operators, where $\mathcal{A}_\Omega$ contains disorder operators and $\mathcal{A}_{\text{phys}}$ contains purely physical operators. For simplicity, we identify $\mathcal{A}_\Omega$ and $\mathcal{A}_{\text{phys}}$ with their canonical inclusions into $\mathcal{A}$. Each disorder operator in $\mathcal{A}_\Omega$ commutes with every element of $\mathcal{A}$.

A disorder-averaged ground state is a linear map $\braket{\cdot} : \mathcal{A} \to \mathbb{C}$ which satisfies
\begin{align}
& \text{Disorder expectations: }\braket{R} = \mathbb{E}[R]\label{eq:general-moments}\\
& \text{Positivity: }\braket{\mathcal{O}^\dagger \mathcal{O}} \geq 0\label{eq:general-positivity}\\
& \text{Perturbative Positivity: }\braket{\mathcal{O}^\dagger [H, \mathcal{O}]} \geq 0\label{eq:general-perturbative-positivity}
\end{align}
for all $R \in \mathcal{A}_\Omega$ and $\mathcal{O} \in \mathcal{A}$, where $\mathbb{E}[R]$ is the expectation value of $R$ with respect to the disorder probability distribution. The Hamiltonian $H$ is not required to be an operator, rather it is defined through its commutator action on $\mathcal{A}$. In particular, $[H, R] = 0$ for $R \in \mathcal{A}_\Omega$. \cref{eq:general-moments} specifies the probability distribution through its moments, \cref{eq:general-positivity} selects for physical states, and \cref{eq:general-perturbative-positivity} selects specifically for ground states at each disorder realization.

Similar to the clean case in which ground-state bounds are obtained by optimizing an expectation value subject to normalization, positivity, and perturbative positivity, we can now bootstrap disorder-averaged ground-state bounds by optimizing an unquenched expectation value subject to \cref{eq:general-moments,eq:general-positivity,eq:general-perturbative-positivity}. In order to produce finite optimization problems, we enforce \cref{eq:general-moments} only for $R$ in a finite set $\mathcal{R} \subset \mathcal{A}_\Omega$ and \cref{eq:general-positivity,eq:general-perturbative-positivity} only for $\mathcal{O}$ in the complex linear span of a finite set $\mathcal{P} \subset \mathcal{A}$. While we are interested in ground-state observables in this study, we note that it is possible to generalize this construction to thermal states by replacing perturbative positivity with the energy-entropy balance inequality \cite{Fawzi_2024,Cho:2025vws}.

In the case of a two-replica system, the algebra of quasi-local operators is $\mathcal{A}^{(2)} = \mathcal{A}_\Omega \otimes \mathcal{A}_{\text{phys}} \otimes \mathcal{A}_{\text{phys}}$. We define linear inclusion maps $r_j : \mathcal{A} \to \mathcal{A}^{(2)}$ for $j = 1, 2$ generated by $r_1(R \otimes P) = R \otimes P \otimes \mathbb{1}_{\text{phys}}$ and $r_2(R \otimes P) = R \otimes \mathbb{1}_{\text{phys}} \otimes P$ for $R \in \mathcal{A}_\Omega$ and $P \in \mathcal{A}_{\text{phys}}$, where $\mathbb{1}_{\text{phys}} \in \mathcal{A}_{\text{phys}}$ is the identity operator. We then have
\begin{equation}\label{eq:general-PPT}
\text{PPT: } \braket{r_1(\mathcal{O})^\dagger r_2(\mathcal{O})} \geq 0
\end{equation}
for all $\mathcal{O} \in \mathcal{A}$. To produce finite optimization problems, we enforce \cref{eq:general-moments} only for $R$ in a finite set $\mathcal{R} \subset \mathcal{A}_\Omega$, we enforce \cref{eq:general-positivity,eq:general-perturbative-positivity} only for $\mathcal{O}$ in the complex linear span of a finite set $\mathcal{P} \subset \mathcal{A}^{(2)}$, and we enforce \cref{eq:general-PPT} only for single-replica operators $\mathcal{O} \in \mathcal{P} \cap \mathcal{A}$. Further details of the two-replica case are discussed in \cref{sec:replica}. Going forward, we use the simplified notation introduced here.

\begin{figure*}
    \centering
    \includegraphics[width=\linewidth]
        {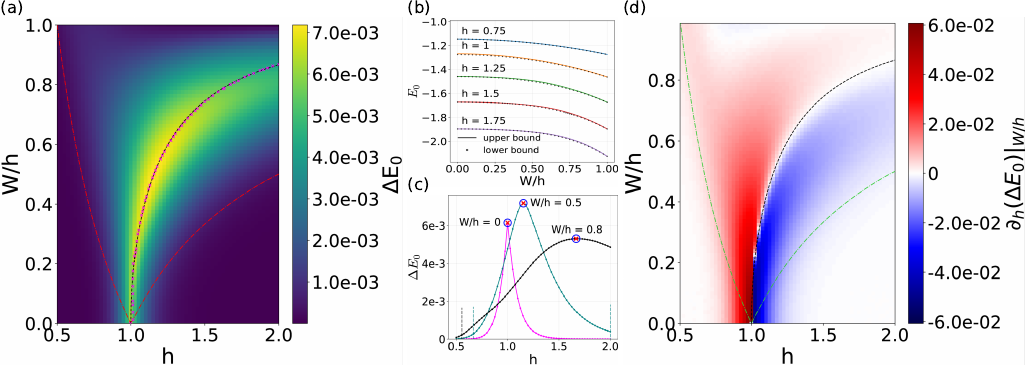}

    \caption{{\bf Bounds on disorder-averaged ground-state energy density for $\mathbb{Z}_2$ disorder.} (a) Colormap of the difference between upper and lower bootstrap bounds (the bootstrap gap $\Delta E_0$), as a function of relative disorder strength $W/h$ and field $h$. The red lines indicate the crossover to the Griffiths region and the black line indicates the analytical critical point. Zeros of the partial derivative $\partial_h \Delta E_0|_{W/h}$ of the bootstrap gap with respect to $h$ at fixed $W/h$ are indicated with magenta crosses. (b) Upper and lower bootstrap bounds on the ground-state energy density for the RTFIM with $\mathbb{Z}_2$ disorder for $h=0.5,1,1.5,2$. The maximum bootstrap gap over the entire scanned set is less than $0.0072$. (c) Bootstrap gap plotted with respect to $h$ for $W/h=0,0.5,0.8$. The analytical critical point is indicated with a red cross and the data points with maximum bootstrap gap are indicated with the blue circle. The crossover to the Griffiths region is indicated with vertical lines. (d) Colormap of the partial derivative $\partial_h \Delta E_0|_{W/h}$. Green lines indicate the crossover to the Griffiths region and the black line indicates the critical line.}
    \label{fig:energy_z2_bounds_combined}
\end{figure*}

\textit{Model.}---We consider the infinite 1D RTFIM with Hamiltonian
\begin{equation}
H = -\sum_j Z_j Z_{j+1} - (h + W\xi_j) X_j. \label{eqn:rtfimspin}
\end{equation}
Here, $X_j$, $Y_j$, and $Z_j$ are Pauli operators for site $j$, the $\xi_j$ are independent and identically distributed (iid) real-valued RVs, and $h$ and $W$ are real parameters. Independence implies that disorder expectation values factorize between sites:
\begin{equation}
\mathbb{E}[\xi_j^{n_j} \xi_k^{n_k}] = \mathbb{E}[\xi_j^{n_j}] \mathbb{E}[\xi_k^{n_k}] \text{ for } j \neq k.\label{eqn:disorderindependent}
\end{equation}
Noting that the probability distribution is determined by the distribution of $\xi_0$, we consider the following three cases:
\begin{enumerate}
\item ($\mathbb{Z}_2$) $\xi_0$ is a discrete RV taking values $1$ and $-1$, each with probability $\frac{1}{2}$.
\item (Uniform) $\xi_0$ is a uniform RV on the interval $[-1,1]$.
\item (Gaussian) $\xi_0$ is a Gaussian RV with mean $0$ and standard deviation $1$.
\end{enumerate}

The Hamiltonian in \cref{eqn:rtfimspin} can be rewritten using Majorana operators through the Jordan-Wigner (JW) transformation:
\begin{gather}
X_j  = -i\gamma_{2j-1}\gamma_{2j}\qquad
Y_j  = \prod_{m<j}(-i\gamma_{2m-1}\gamma_{2m} )\gamma_{2j} \nonumber \\
Z_j  = \prod_{m<j} (-i\gamma_{2m-1}\gamma_{2m}) \gamma_{2j-1}.
\label{eq:jordan-wigner}
\end{gather}
This yields
\begin{equation}
H = i\sum_j \gamma_{2j}\gamma_{2j+1} + (h+W\xi_j) \gamma_{2j-1}\gamma_{2j}.
\end{equation}
We use the Majorana representation so that in the clean case we obtain a free fermion Hamiltonian which can be solved very accurately with bootstrap \cite{berenstein2024numericalexplorationbootstrapspin,hastings2024perturbationtheorysumsquares,Scheer_2026}. This avoids the difficulties associated with domain walls discussed in Refs. \cite{scheer2025defectbootstraptightground, chadha2026bootstrapboundsquantumspin}.

In the bootstrap program, we take $\mathcal{P}$ to be a set of monomials of the form 
\begin{equation}\label{eq:define-monomial}
\begin{split}
&\xi_{j_0} \cdots \xi_{j_n} \gamma_{k_0} \cdots \gamma_{k_m} \text{ with }\\
&j_0 \leq \cdots \leq j_n \text{ and } k_0 \leq \cdots \leq k_m.
\end{split}
\end{equation}
Products and adjoints of these monomials can be simplified using the relations
\begin{equation}
\begin{split}\label{eq:algebraic-relations}
\{\gamma_j, \gamma_k\} &= \delta_{j,k}, \quad [\xi_j, \xi_k] = 0, \quad [\xi_j, \gamma_k] = 0\\
\gamma^\dagger_j &= \gamma_j, \quad \xi^\dagger_j = \xi_j.
\end{split}
\end{equation}
In the case of $\mathbb{Z}_2$ disorder, we additionally use the relation
\begin{equation}\label{eq:xi-squared}
\xi_j^2 = 1.
\end{equation}
The set $\mathcal{R}$ consists of monomials of the form $\xi_{j_0} \cdots \xi_{j_n}$. The specific sets $\mathcal{P}$ and $\mathcal{R}$ are described in \cref{sec:numerics}.

The critical properties of this model can be calculated using strong-disorder RG \cite{PhysRevLett.69.534}. The critical point is characterized by $\mathbb{E}[\log |h + W\xi_0|] = 0$. Additionally, Griffiths effects are seen when rare regions within the Ising chain can exhibit order that is characteristic of the phase opposite to that of the bulk phase. This happens when the disorder is strong enough to allow local couplings to enter the opposite phase. Thus, the ferromagnetic Griffiths region is defined by $\mathbb{E}[\log| h + W\xi_0|] < 0, \max(h + W\xi_0) > 1$ and the paramagnetic Griffiths region is defined by $\mathbb{E}[\log |h + W\xi_0|] > 0, \min(h + W\xi_0) < 1$. 

\textit{Results.}---Bootstrap lower and upper bounds for the disorder-averaged ground-state energy density $E_0$ with $\mathbb{Z}_2$ disorder are shown in \cref{fig:energy_z2_bounds_combined}, where we scan over the range $h \in [0.5,2]$, $W/h \in [0,1]$\footnote{We note that all couplings $\xi_i$ can be made non-negative by a simple site-local unitary transformation. This is the reason for the specific choice of $W/h$.}. The bounds are tight throughout the phase diagram; the maximum difference between upper and lower bounds on the ground-state energy density (the bootstrap gap) is less than $0.0072$.

While the bootstrap gap is small throughout the phase diagram, a remarkable observation is that it
provides meaningful information about the phase diagram of the RTFIM. The colormap of the bootstrap gap plotted in \cref{fig:energy_z2_bounds_combined}(a) demonstrates its behavior throughout the phase diagram. As a function of $h$ at fixed disorder strength, the bootstrap gap is small and flat in the weakly-disordered paramagnetic and ferromagnetic phases, indicated by the dark blue regions in the diagram. The bootstrap gap begins increasing near the Griffiths lines, and peaks near the critical points indicated by the yellow region.

\begin{figure}[t]
    \centering
    \includegraphics[width=\linewidth]{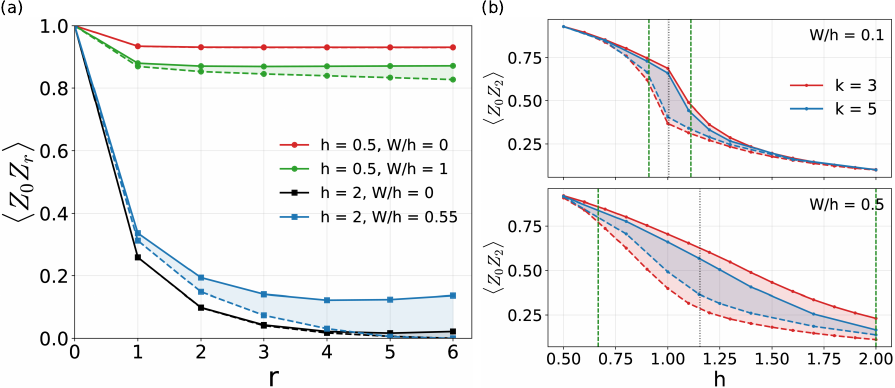}
    \caption{{\bf Bounds on disorder-averaged ground-state spin-spin correlators $\braket{Z_0 Z_r}$ for $\mathbb{Z}_2$ disorder.} (a) $\braket{Z_0 Z_r}$ as a function of $r$ for weakly and strongly disordered ferromagnetic and paramagnetic parameters. (b) $\braket{Z_0 Z_2}$ as a function of $h$ for $W/h=0.1,0.5$. Results are shown for two different operator sets with Krylov depth $k=3,5$ (see \cref{tab:observable_sets}). The vertical green lines indicate the onset of Griffiths regions and the vertical black lines indicate the critical point.}
    \label{fig:zozr_bounds_combined}
\end{figure}

To better highlight the bootstrap gap's ability to identify phase boundaries, we evaluate $\partial_h \Delta E_0|_{W/h}$ where $\Delta E_0$ is the bootstrap gap and the derivative with respect to $h$ is taken at fixed relative disorder $W/h$. Rather than obtaining this quantity from the energy bounds directly through finite differences, we instead use the SDP gradient formula based on the envelope theorem \cite{Reehorst_2021} which allows us to directly access this derivative. The line of zeros of $\partial_h \Delta E_0|_{W/h}$, indicated by magenta crosses, is nearly coincident with the analytical critical line. The colormap of the partial derivative, presented in \cref{fig:energy_z2_bounds_combined}(d), shows the weakly-disordered regions in white, where the bootstrap gap is small and flat. Additionally, we see that the partial derivative changes its sign across the critical line.
\begin{figure}
    \centering
    \begin{subfigure}[b]{0.8\linewidth}
    \centering
    \includegraphics[width=\linewidth]{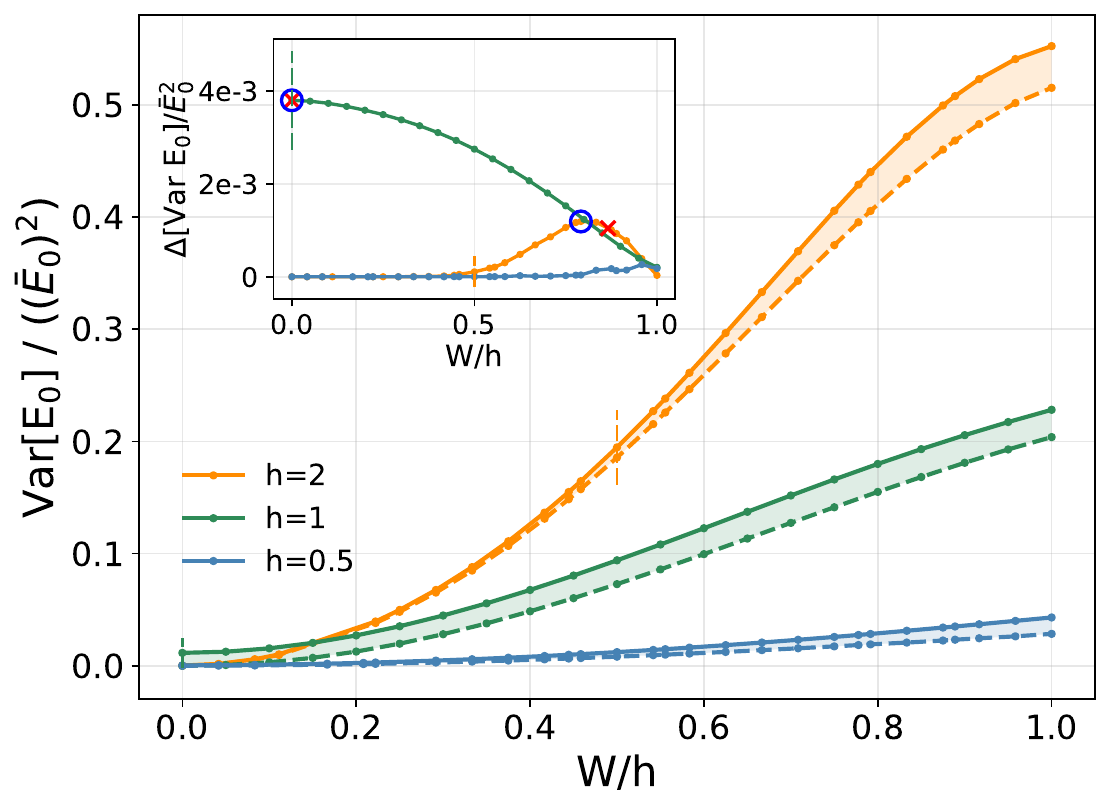}
    \end{subfigure}
    \caption{{\bf Bounds on the scaled energy density variance for $\mathbb{Z}_2$ disorder.} The variance is scaled by the square of the average of lower and upper bounds on the energy density, with the differences between upper and lower bounds shown in the inset. The onset of the Griffiths region is indicated with vertical lines. The maximum of the difference between upper and lower bounds (indicated with blue circles in the inset) is maximized near the critical point (indicated with red crosses in the inset).}\label{fig:energy_variance_bounds_combined}
\end{figure}
\begin{figure*}
    \centering \includegraphics[width=0.8\linewidth]{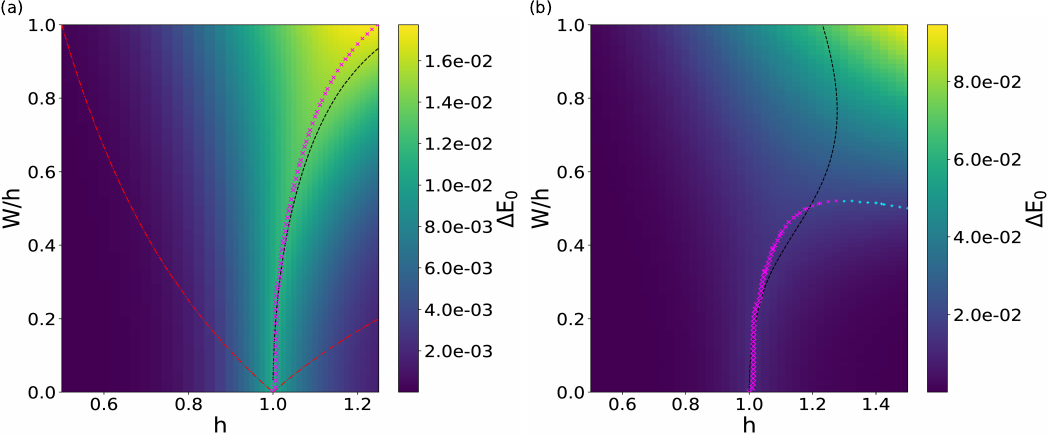}
    \caption{{\bf Bootstrap gap for uniform and Gaussian disorder.} Colormap of difference between upper and lower bootstrap bounds on ground-state energy density (the bootstrap gap $\Delta E_0$) of the RTFIM with (a) uniform and (b) Gaussian disorder. The Griffiths lines for uniform disorder are indicated in red and the analytical critical line is indicated in black. The magenta crosses indicate zeros for the partial derivatives with respect to $h$ of the bootstrap bounds at fixed $W/h$. The blue plus signs in (b) indicate inflection points in the bootstrap gap.}
    \label{fig:energy_continuous_bounds_combined}
\end{figure*}
Bounds on the disorder-averaged ground-state spin-spin correlators $\langle Z_0Z_r\rangle$ are evaluated up to $r=6$ using the identity
\begin{equation}\label{eq:Z0-Zr}
Z_0 Z_r=(-i)^{r}\prod_{0\leq j<2r-1}\gamma_j.
\end{equation}
Bounds as a function of $r$ are shown in \cref{fig:zozr_bounds_combined}(a) for weak and strong disorder in the paramagnetic and ferromagnetic phases. In the paramagnetic phase at large $r$, the bootstrap lower bound on the spin-spin correlator may be negative. This can be fixed by incorporating a subset of Griffiths inequalities \cite{Miyao_2016} in the bootstrap program when bounding large-$r$ spin-spin correlators. See \cref{sec:numerics} for more details.

In \cref{fig:zozr_bounds_combined}(b), we plot bounds on $\braket{Z_0Z_2}$ for $W/h = 0.1,0.5$ as a function of $h$. Similarly to the case of the energy density, we find that the difference between the upper and lower bounds on spin-spin correlators starts to increase significantly in the Griffiths region (indicated by green lines) and peak near the critical points (indicated by black lines).

We emphasize that while our bounds apply in the thermodynamic limit, the bootstrap gap strongly depends on the size and support of the operator set $\mathcal{P}$. This is already evident in \cref{fig:zozr_bounds_combined}(b) where the bounds get tighter as we increase the number of Krylov steps, which increases the size and support of $\mathcal{P}$ (see \cref{sec:numerics}). To understand why the bootstrap gap displays features similar to critical scaling, we make the following observation. By translation symmetry (see \cref{sec:numerics}), the bootstrap constraints for a set $\mathcal{P}$ containing operators with support in a region of size $L$ are valid for a finite system of size at least $L+1$ with periodic boundary conditions. Therefore, the bootstrap gap has a rigorous interpretation as an upper bound on the energy density difference $\mathcal{E}_{L+1}$ between a length $L+1$ periodic system and an infinite size system. Since $\mathcal{E}_L$ is a local expectation value, its convergence to $0$ is generally power law in $L$ at criticality or in a Griffiths region and exponential in $L$ in a gapped phase \cite{Cardy_1996,PhysRevLett.69.534,Hastings_2006}. Although the bootstrap gap using our Krylov operator sets $\mathcal{P}$ does not necessarily saturate this bound, we nonetheless observe the power law to exponential crossover at the Griffiths lines. This is analyzed in more detail in \cref{sec:krylov}, \cref{fig:krylov_analysis}, where we find exponential convergence in the Krylov-depth throughout the weakly disordered phases and power law convergence within the Griffiths regions and near the critical line, with the crossover between the two closely tracking the Griffiths lines.

Next, we discuss bounds on the variance of the ground-state energy density with respect to disorder, defined as $\text{Var} E_0 =\langle H_0^2 \rangle-\langle H_0\rangle^2$. This is computed using the two-replica bootstrap program. We rescale the variance bounds by the square of the mean of the upper and lower bootstrap energy density bounds to obtain a value in the interval $[0,1]$. The bounds are plotted in \cref{fig:energy_variance_bounds_combined} for $h = 0.5,1,2$ with the difference between upper and lower bounds shown in the inset. The overall scale of the variance increases as the magnitude of the disordered on-site transverse field term is increased. Furthermore, we find that the difference between the upper and lower bootstrap bounds on the rescaled variance is maximized near the critical point.

Results analogous to those obtained above for the $\mathbb{Z}_2$ disorder case can be straightforwardly obtained for continuous disorder distributions as well. In this case, we cannot simplify algebraic expressions using \cref{eq:xi-squared} so our program becomes more costly for a fixed Krylov depth. We find that the zeros of  $\partial_h \Delta E_0|_{W/h}$ provide a good approximation to the analytical critical line up to a certain value of $W/h$ (see \cref{fig:energy_continuous_bounds_combined}). More detailed results on continuous disorder distributions including bounds, bootstrap gaps, and partial derivative colormaps are provided in \cref{sec:detailedcontinuous}.

\textit{Discussion.}---We have presented a bootstrap framework for quantum systems with quenched disorder based on an extension of the operator algebra to include classical disorder variables. The resulting bounds are rigorous, two-sided, valid directly in the thermodynamic limit, require no sampling of disorder realizations, and can access multi-replica correlators. For the RTFIM, we obtain tight bounds on the disorder-averaged ground-state energy density, short-range spin-spin correlators, and disorder variance of the energy density. One of our main observations is that the bootstrap gap (the difference between upper and lower disorder-averaged ground-state energy density bounds) is a useful diagnostic of the phase. Since the constraints generated by an operator set of finite range are also satisfied by sufficiently large finite systems, the bootstrap gap upper bounds the finite-size correction to the energy density and inherits its scaling: it scales exponentially in the support of the operator set in the weakly disordered phases, crosses over to a power law scaling at the onset of Griffiths region, and peaks along the critical line. Although we focused on the energy density bounds here, this idea can also be applied to other local expectation values. Additionally, while we focused on RTFIM, our approach is not tied to one dimension or to free fermions and can be generalized to higher dimensions and more complicated interacting models with disorder and finite temperature.

\textit{Acknowledgements.}---We thank Akshat Panday, Pavel A. Nosov, and Zhaoyu Han for valuable conversations. MGS is supported by a postdoctoral fellowship from the Julian Schwinger Foundation. The work of MC is supported by the Simons Collaboration on Global Categorical Symmetries. Research reported in this publication was supported by an award from the Harvard FAS Dean’s Competitive Fund for Promising Scholarship. The computations in this paper were performed using the FASRC Cannon cluster, supported by the FAS Division of Science Research Computing Group at Harvard University. 

\bibliography{references}

\appendix
\section{Replica Observables\label{sec:replica}}
We now give further details regarding the bootstrap for two-replica systems. Since two-replica programs require significantly more resources than single-replica programs in order to achieve the equivalent accuracy, we include bounds derived from the single-replica program as inequality constraints in the two-replica program to improve results. Specifically, in our energy variance computation, we impose as inequality constraints the upper and lower energy density bounds we derive using the single-replica bootstrap.

The variance of a single-replica observable $\mathcal{O} \in \mathcal{A}$ is given by
\begin{equation}
\text{Var}[\mathcal{O}] = \braket{r_1(\mathcal{O}) r_2(\mathcal{O})} -\frac{1}{4}(\braket{r_1(\mathcal{O})} + \braket{r_2(\mathcal{O})})^2.\label{eqn:variance}
\end{equation}
The expression in \cref{eqn:variance} can be lower bounded by using the two-replica bootstrap to derive a lower bound on the first term and the one-replica bootstrap to derive an upper bound on the second term. A combination of these bounds can be used to derive a lower bound on the variance. The disadvantage of this approach is that the lower bound inherits slack from two independent programs. For observables whose bounds are loose, this approach may not lead to a bound stronger than the trivial zero lower bound on the variance imposed by positivity.

On the other hand, an upper bound on the variance can be computed directly within a single instance of the replica bootstrap by noting that
\begin{equation}\label{eq:variance-condition}
\begin{pmatrix}
\langle r_1(\mathcal{O})r_2( \mathcal{O})\rangle-t  &  \frac{1}{2}(\braket{r_1(\mathcal{O})} + \braket{r_2(\mathcal{O})})  \\
\frac{1}{2}(\braket{r_1(\mathcal{O})} + \braket{r_2(\mathcal{O})})  & 1 
\end{pmatrix} \succeq 0
\end{equation}
implies $\text{Var}[\mathcal{O}] \geq t$. We can implement the condition in \cref{eq:variance-condition} by adding the auxiliary variable $t$ into the optimization problem. A lower bound on the set of $t$ for which \cref{eq:variance-condition} is feasible constitutes an upper bound on $\text{Var}[\mathcal{O}]$.

\section{Implementation Details\label{sec:numerics}}
\begin{table*}[htbp]
  \centering
  \begin{tabular}{lccc}
    \hline\hline
    Observable & $\mathcal{P}$ $(d,k)$ & $|\mathcal{P}|$ & Median solve time \\
    \hline
    Z$_2$ energy    & $(3,8)$                                       & $731$            & $\sim\!10.6$ min \\
    Continuous Energy  & $(2,6)$                                       & $695$            & $\sim\!24.1$ min \\
    Energy variance & $(1,4)$ per replica & $1299$ & $\sim\!37.2$ min \\
    $Z_0Z_r$        & $(3,5)$, $+\,Z_0Z_2,\,Z_0Z_3$        & $1927$           & $\sim\!10.3$ h   \\
    \hline\hline
  \end{tabular}
  \caption{Operator sets used in the bootstrap computations. The seed set $\mathcal{S}_0$ contains operators $\xi_j$ and $\gamma_j$ for $-d \leq j \leq d$. We denote the number of Krylov steps by $k$ so that $\mathcal{P} = \mathcal{S}_k$. For calculations of the $Z_0Z_r$ correlators, the $Z_0Z_2$ and $Z_0Z_3$ Jordan-Wigner string monomials are additionally included in $\mathcal{S}_0$ so the Krylov iteration generates larger distance $Z$ products. For the two-replica case, for each operator of the form $R \otimes P$ in the one-replica seed set, where $R\in A_\Omega$ and $P\in\mathcal{A}_{\text{phys}}$, we include $R \otimes P \otimes \mathbb{1}_{\text{phys}}$, $R \otimes \mathbb{1}_{\text{phys}} \otimes P$, and $R \otimes P \otimes P$ in the two-replica seed set.}
  \label{tab:observable_sets}
\end{table*}

Here we provide further details on the numerical implementation including our choices of $\mathcal{P}$ and $\mathcal{M}$, the difference between discrete and continuous disorder distributions, and the role of symmetries.

We iteratively construct an operator set $\mathcal{P}$ by starting with an initial local seed basis $\mathcal{S}_0$ of monomials of the form in \cref{eq:define-monomial}. For each $k \geq 1$, we construct $\mathcal{S}_k$ from $\mathcal{S}_{k-1}$ by the following procedure. For each $p \in S_{k-1}$, we decompose $[H, p]$ into a linear combination of monomials of the form in \cref{eq:define-monomial} using \cref{eq:algebraic-relations} and in the case of $\mathbb{Z}_2$ disorder also \cref{eq:xi-squared}. We add all of the resulting monomials to $S_k$. We refer to each of these steps as a ``Krylov step" because it involves commutation with $H$. Empirically, we find that taking $\mathcal{P} = \mathcal{S}_k$ produces programs with better performance than many other constructions. The specific operator sets used in the bootstrap program are listed in \cref{tab:observable_sets}.

We write the bootstrap constraints as a semidefinite program (SDP) by constructing a set $\mathcal{Q}$ of monomials of the form in \cref{eq:define-monomial} whose span contains all operators of the form $p_1^\dagger p_2$ or $p_1^\dagger [H,p_2]$ for $p_1, p_2 \in \mathcal{P}$. \cref{eq:general-positivity,eq:general-perturbative-positivity} can then be expressed as positive semidefinite constraints as explained in Ref. \cite{Scheer_2026}. We impose \cref{eq:general-moments} for operators $R$ in $\mathcal{R} = \mathcal{Q}\cap \mathcal{A}_\Omega$. In addition to the independence constraints listed in \cref{eqn:disorderindependent}, we require disorder moments for the three distributions. We have
\begin{equation}
\begin{split}
\text{$\mathbb{Z}_2$: }\mathbb E [\zeta_j] &= 0 \\
\text{Uniform: }\mathbb E[\zeta_j^n] &= \begin{cases} 
\frac{1}{n+1} & n \text{ is even}\\
0 & n \text{ is odd}
\end{cases}\\
\text{Gaussian: }\mathbb{E}[\zeta_j^n] & = \begin{cases}
(n-1)!! & n \text{ is even}\\
0 & n \text{ is odd}.
\end{cases}
\end{split}
\end{equation}
For $\mathbb{Z}_2$ disorder, moments higher than first-order are simplified away using \cref{eq:xi-squared}.

Symmetries that only hold on average can be realized as standard unitary or antiunitary symmetries of the unquenched Hamiltonian. For example, in the RTFIM, translation symmetry is given by
\begin{equation}
T \gamma_j T^\dagger = \gamma_{j+2},\quad T \xi_j T^\dagger = \xi_{j+1}.
\end{equation}
In addition to translation symmetry, the RTFIM has reflection, number parity, and complex conjugation symmetries. We use these symmetries to reduce the number of constraints and variables in the SDP using the methods of Ref. \cite{Scheer_2026}. For two-replica systems, in addition to the joint action of the individual symmetries on each replica, replica-swap symmetry is also included.

In order to obtain useful lower bounds on long-range spin-spin correlators, we impose the Griffiths inequalities (detailed in Ref. \cite{Miyao_2016}) up to second-order in spin-spin correlators:
\begin{align}
\langle Z_0Z_r\rangle \geq 0,\quad \langle(1+X_0)(1+X_r)\rangle \geq 0.
\end{align}
These inequalities are imposed for all pairwise products contained in $\mathcal{Q}$ and are written in the Majorana basis using the JW transform in \cref{eq:jordan-wigner}. The Griffiths inequalities are omitted from bootstrap programs bounding energy density and energy variance because their influence on these bounds is negligible.

Semidefinite programs were solved using MOSEK \cite{mosek}. We acknowledge the use of Claude Opus 5 (Anthropic) as an assistant for code development. All code was reviewed and tested by the authors.

\section{Uniform and Gaussian Disorder\label{sec:detailedcontinuous}}
\begin{figure*}[t]
    \centering
        \centering
        \includegraphics[width=\linewidth]{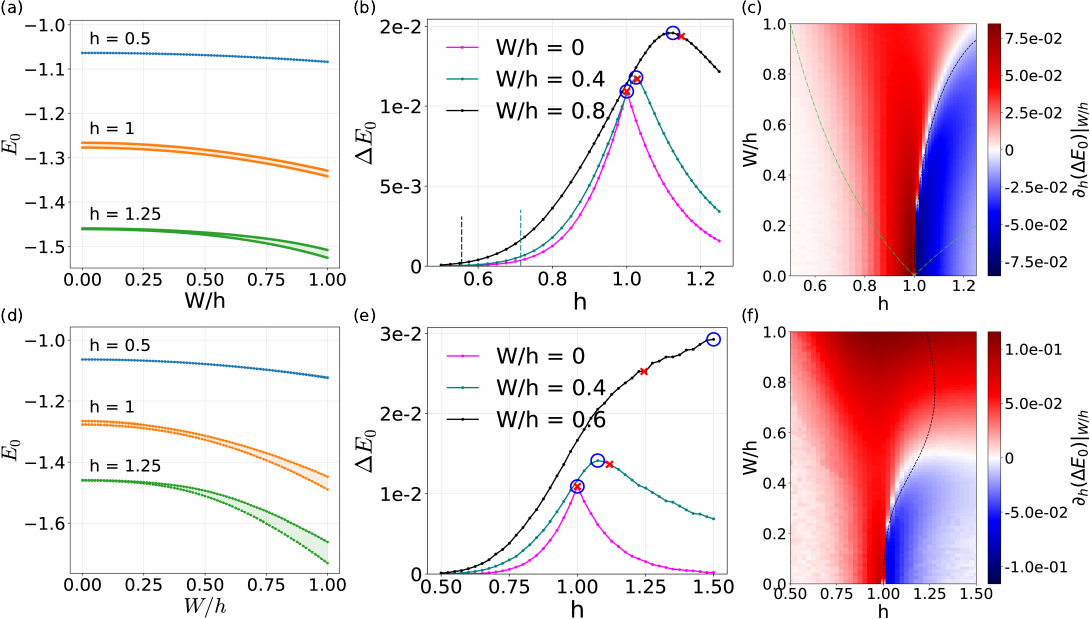}
    \caption{{\bf Bounds on disorder-averaged ground-state energy for uniform (top row) and Gaussian (bottom row) disorder.} From left to right: (a,d) Upper and lower bootstrap bounds for uniform disorder and $h=0.5,1,1.25.$, (b,e) bootstrap gap for $W/h=0,0.4,0.6$, with the analytical critical point is shown with a red cross and the zeros of $ \partial_h \Delta E_0|_{W/h}$ indicated with the blue circles. The crossover to the Griffiths region is indicated with vertical lines. (c,f) colormap of $ \partial_h \Delta E_0|_{W/h}$, with Griffiths lines indicated in green and the critical line indicated in black, demonstrating the sign change near the critical line. }
        \label{fig:continuous_disorder_combined}
\end{figure*}
Here we provide further results on bootstrap bounds obtained for the RTFIM with continuous disorder distributions. In \cref{fig:continuous_disorder_combined}, we provide results on upper and lower bootstrap bounds on the ground-state energy density, the bootstrap gap, and the partial derivative $ \partial_h \Delta E_0|_{W/h}$. Compared to the $\mathbb{Z}_2$ case, the bounds for the continuous disorder distributions are looser, in part because we cannot simplify expressions using \cref{eq:xi-squared}. The signature for the Griffiths region is less clean for uniform disorder than for $\mathbb{Z}_2$ disorder, because the entrance into the Griffiths region is more gradual for a uniform distribution. The Gaussian contains no Griffiths boundaries because its support is unbounded. However, the color map of the partial derivative in \cref{fig:continuous_disorder_combined}(f) still contains white regions near small disorder indicating flat bounds characteristic of the weakly disordered phase, indicating that Griffiths effects may still occur.

The zeros of the partial derivative track the analytical critical line well for $W/h\lesssim 0.6$ for uniform disorder and $W/h \lesssim 0.3$ for Gaussian disorder. It is worth noting that while the $\mathbb{Z}_2$ and uniform cases exhibit peaks in the fixed-$W/h$ vs. $h$ diagrams throughout all scanned couplings (see \cref{fig:energy_z2_bounds_combined}(b) and \cref{fig:energy_continuous_bounds_combined}(b)), the equivalent diagram for Gaussian disorder fails to show this peak behavior for $W/h\gtrsim 0.5$, with bootstrap gap steadily increasing with $h$ after that point. Indeed, the white regions in \cref{fig:energy_continuous_bounds_combined}(f) for $h \gtrsim 1.2$ are bootstrap gap inflection points rather than maxima. This is close to the region where the critical line starts to slope downward in $h$ due to the decreasing expectation value of $\mathbb E[\log \xi_0]$. We attribute this to the fact that the moments entering the bootstrap program provide limited use in pinning down the physical disorder distribution in this regime, and so the bootstrap proxy for criticality fails to accurately track the true critical line in this regime.

\section{Convergence of Bootstrap Gap with Krylov Depth\label{sec:krylov}}
\begin{figure*}[t]
    \centering
        \centering
        \includegraphics[width=\linewidth]{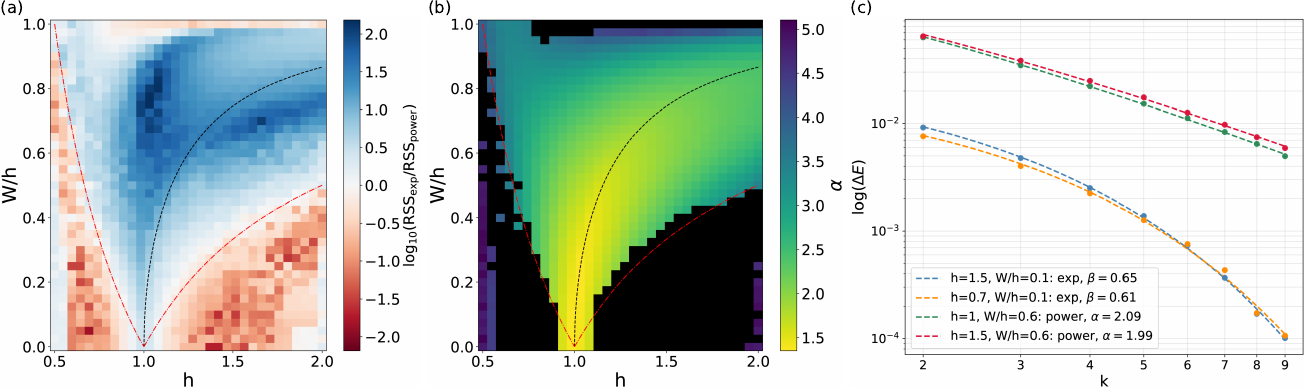}
    \caption{{\bf Analysis of Krylov convergence of energy density bootstrap gap.} Red lines indicate the onset of the Griffiths regions and the critical line is shown in black. (a) Colormap of relative RSS errors of the exponential fit against the power-law fit, computed as $\log \text{RSS}_{\text{exp}}/\text{RSS}_{\text{power}}$. (b) Colormap indicating choice of power-law or exponential fit to bootstrap gap scaling with Krylov depth. Black regions indicate where the exponential fit returns lesser RSS error. Colored regions indicate a power-law scaling with exponent $\alpha$.  The red regions indicate an exponential fit whereas blue regions indicate a power-law. (c) Representative fits to power-law and exponential convergence in Krylov depth, shown for $W/h = 0.1$ $h =0.7,1.5$ and $W/h = 0.6$, $h=1,1.5$.}
        \label{fig:krylov_analysis}
\end{figure*}
To investigate the dependence of the bootstrap gap on the size and support of the operator set, we evaluate the evolution of the bootstrap gap with increasing Krylov step for a simple seed basis with $d=1$ and $k\in[2,9]$, using the notation in \cref{tab:observable_sets}. We fit the bootstrap gap $\Delta E_0$ as a function of $k$ to exponential and power-law forms, $\log \Delta E_0 = A- \alpha \log k$ or $\log \Delta E_0 = A - \beta k$. We may then interpret $\alpha$ as a dynamical scaling exponent and $\beta$ with the inverse correlation length. \cref{fig:krylov_analysis}(a) shows a colormap of the ratio of the residual-sum-of-squares (RSS) errors of the exponential and power-law fits on a logarithmic scale. 
In \cref{fig:krylov_analysis}(b), the black regions indicate the points where the exponential fit has lower RSS error. At all other points, the colormap shows the power-law exponent $\alpha$. In \cref{fig:krylov_analysis}(b), the red lines indicating the crossover into the Griffiths phase coincide with a switch from exponential to power-law convergence of the bootstrap gap. We also note that the power-law exponent near the critical line is approximately $\alpha = 2$, which is the expected value from a clean free-fermion model.

We note that the evolution of the bootstrap gap as a function of Krylov step is not a perfect description of the finite-size scaling of this quantity because the operator set $\mathcal{P}$ contains only a subset of the operators which occur in a given region. In the companion work on classical disordered bootstrap \cite{unpubl1}, a scaling analysis was performed on the 1D disordered contact process using a bootstrap program which includes all operators in a given region. The dynamical scaling exponent $\alpha=1$ was recovered accurately. A more detailed analysis of the dependence of the bootstrap gap on the complexity, diameter, and maximum support of the operator set is warranted for a more careful interpretation of the bounds scaling with Krylov step.

\end{document}